\documentstyle[prd,aps,epsf]{revtex}

\begin{document}


\title{Gravitational instantons and internal dimensions.}

\author{E. J. Copeland$^{1}$\thanks{E-mail:
edmundjc@pcss.maps.susx.ac.uk},
        J. Gray$^{1}$\thanks{E-mail:
kapm3@pcss.maps.susx.ac.uk}\&,
        P. M. Saffin$^{2}$\thanks{E-mail:
p.m.saffin@damtp.cam.ac.uk}}
\address{
$^1$Centre for Theoretical Physics, University of Sussex,
Falmer, Brighton.\\
$^2$DAMTP, CMS, Wilberforce Rd, Cambridge, CB3 0WA.
}
\date{\today}
\maketitle

\begin{abstract}
We study instanton solutions in general relativity with a scalar field.
The metric ansatz we use is
composed of a particular warp product of general 
Einstein metrics, such as those found in
a number of cosmological settings, including string
cosmology, supergravity compactifications and general Kaluza Klein 
reductions. Using the Hartle-Hawking prescription the
instantons we obtain determine whether metrics involving extra compact
dimensions of this type are favoured as initial conditions for the
universe. Specifically, we find that these product metric instantons,
viewed as constrained instantons,
do have a local minima in the action. These minima are then
compared with the higher dimensional version of the
Hawking-Turok instantons, and we argue that the latter always have
lower action than those associated with these product metrics.
\end{abstract}

\pagestyle{myheadings}
\thispagestyle{empty}
\markright{
\hspace{7em}
\small Copeland, Gray and Saffin --
\emph{Gravitational instantons and internal dimensions.}
}



\section{Introduction}
Consistent unified field theories which include gravity appear to
indicate that the Universe has more than four spacetime
dimensions. An obvious problem which follows is how to
interpret these unseen extra dimensions? One approach that has
been followed is to postulate that only four of these are
observable, the extra dimensions have managed 
to become compact and are
unobservably small. Recently however there has been a tremendous
amount of interest in the effective five dimensional cosmologies
associated with Branes, in which the fifth dimension can
be macroscopic in size, yet remain unobservable at low energies. In
general, these compactified spaces are assumed as part of the
initial metric ansatz, and the cosmology of such metrics is then
determined. Although this is a natural approach to take, it does
not address the issue of whether such an initial condition is to
be expected in string or M -theory, for example. Is there any way
in which we can calculate the probability of the Universe
possessing such compact internal dimensions as an initial
condition? It would be of great interest if it could be shown that
quantum cosmology predicts a manifold with compact extra
dimensions as the most likely initial configuration.

Symmetry arguments usually provide a very powerful tool for
determining which instanton solutions should provide the dominant
contribution (i.e. those with lowest Euclidean action) to the Hartle 
Hawking path integral \cite{hartle},
hence providing the most likely background spacetime. An example is
the Hawking-Moss
instanton, involving a scalar field $\phi$ with potential
$V(\phi$)\cite{moss}. Assuming the potential had a stationary
point at some non-zero value they obtained in four spacetime
dimensions an O(5) symmetric instanton solution where $\phi$ is
constant and the Euclidean manifold is a four sphere.

However, Coleman and De Lucia \cite{coleman80} obtained an
instanton solution of lower action with O(4) symmetry which was
non-singular and corresponded to the nucleation of a bubble of
true vacuum in a sea of false vacuum deSitter space. 
It was used in the earliest
versions of open inflation \cite{bucher95}, because the interior
of such a bubble is in fact an open universe.
Hawking and Turok \cite{hawking98a} took these solutions
one step further, dropping
the requirement for non-singular instanton configurations;
they obtained solutions where
the scalar field potential increased monotonically from a single
minimum.
These solutions also allowed for a natural
continuation to an open universe which was inflating. Moreover,
although the instanton solutions themselves were singular their
action was finite. Indeed they demonstrated a family of solutions
which had lower action than the more symmetric O(5) solution!
The notion that the O(4) symmetry of the Hawking-Turok instanton
was responsible for the low action was tested in \cite{saffin99}.
Treating the instanton as a foliation of squashed rather than
round three spheres, it was found that the O(4) instanton 
was the lowest action solution within this family.
In an interesting paper Garriga \cite{garriga98}
proposed a resolution to the problem of having a singularity
in the solution;
singular instantons can arise from
compactifications of regular higher dimensional instantons when
viewed as lower dimensional objects.

In this paper we investigate the nature of instanton solutions for the
largest range of cosmologically relevant
higher dimensional metrics that have been 
studied to date. Our results will
be of relevance for the study of any 
higher dimensional model which involves
compactifications
on Einstein metrics, i.e. models of string cosmology involving 
compactifications on tori, supergravity compactifications on spheres
and string theories where the compactified
dimensions are Calabi-Yau manifolds. 
In particular we will be investigating
instanton solutions arising from the metric ansatz,
\begin{eqnarray}
\label{M_metric}
\rm{ds}^2&=&\rm{d}\xi^2+a_{(1)}^2(\xi)\rm{ds}_{(1)}^2
                       +a_{(2)}^2(\xi)\rm{ds}_{(2)}^2
                       +...+a_{(T)}^2(\xi)\rm{ds}_{(T)}^2.
\end{eqnarray}
The only restriction on the $ds_{(i)}^2$'s is that they
are Einstein metrics on compact manifolds; the Ricci tensor 
is proportional to the
metric. Of the many solutions that exist,
we will see how a class of these instantons may be continued to
a four dimensional inflating universe,
with a number of static extra dimensions.

\indent In general, because of the non-linear nature of the
equations, the solutions for the scale factors $a_{i}$ are
obtained numerically, and from these we can study the action of
the (generically singular) instantons. 
The most important result we
obtain is that the family of singular instantons of this type 
can provide a local minima of the action for non trivial
extra dimensions.
However, it turns out that in all the cases we examined
the
action of these local minima remains {\it higher} than that of the
corresponding higher dimensional Hawking-Turok instanton. The
implication of such a result is important. The symmetry properties
associated with the Hawking-Turok instanton appear to determine
the most likely instanton configuration, at least for the cases
involving Einstein metrics.

\indent The layout of the rest of the paper is as follows: In
section II we derive the field equations and action associated
with our metric. Section III contains the results of our numerical 
and analytical
analysis and presents the nature of the local minima of the
action. It also contains the comparison of these instantons with
the equivalent higher dimensional Hawking-Turok case and shows how
the latter always lead to a lower Euclidean action. Section IV
presents exact
solutions for the case of a cosmological constant replacing the
scalar field potential. We also mention the analytical
continuation of our solutions to a space-time with a lorentzian
signature and demonstrate the existence 
of solutions where the
internal dimensions remain static while the four dimensional
spacetime is inflating. We draw our conclusions in section V.

\section{Derivation of Field Equations}

Our starting point is a manifold ${\cal M}$ which has a metric
structure imposed on it, and a scalar field $\phi$ living on it
with potential ${\cal V}(\phi)$.
By using the usual torsion free, metric connection on ${\cal M}$ we can
describe the equations of motion for the metric and for $\phi$
which follow from the Einstein-Hilbert action.
\begin{eqnarray}
\label{EHaction} S_{\rm{E}}&=&\int_{\cal M} \eta
\left[-\frac{1}{2\kappa^2}{\cal R}+\frac{1}{2}(\partial \phi)^2 +{\cal
V}(\phi)\right]+\textnormal{boundary term}.
\end{eqnarray}
Here, $\kappa^2 = 8\pi /m_{\rm Pl}^2$ (scaled to
unity for the rest of  the paper), and the
boundary terms are such that the action does not
contain second derivatives of the
metric\cite{gibbons77}\cite{barrow89}. $\eta$ is the volume form
and ${\cal R}$ is the Ricci scalar of the connection.

As mentioned earlier we consider the manifold
${\cal M}$ as a foliation in
Euclidean time of a product of boundary-less manifolds. At any
given time $\xi$ we can then write ${\cal M}(\xi)$ as a 
Cartesian product of
${\cal M}_{(i)}$, $i=1...T$ each with dimension $n_{(i)}$, where
for convenience we define \mbox{$N=n_{(1)}+n_{(2)}+...+n_{(T)}$}.
To endow ${\cal M}$ with a metric structure we start by putting a
metric on each of the ${\cal M}_{(i)}$, denoted $\rm{ds}_{(i)}^2$.
The metric structure we impose on ${\cal M}$ then follows by
introducing a $\xi$ dependent scale factor, $a_{(i)}$,
for each ${\cal M}_{(i)}$;
providing information about the relative size of the
${\cal M}_{(i)}$ at any given $\xi$. The resulting metric is then
given by Eq.~(\ref{M_metric}).

This form for the metric is very general. It includes a wide class
of metrics commonly considered in cosmology, such as those leading
to the Coleman-De Lucia instanton \cite{coleman80},
Kantowski-Sachs instantons \cite{jensen89}, Hawking-Turok
instanton \cite{hawking98}, most of the models of string
cosmology arising out of compactifications on tori (for a review
see \cite{lidsey99}), compactifications of string theory on
Calabi-Yau spaces (for a review see \cite{polchinski}), and some
compactifications of Supergravity theories on spheres (for a review
see \cite{duff}). For example in
\cite{coleman80}\cite{hawking98} $T=1$ and ${\cal M}_{(1)}$ is a
three sphere with its standard round metric. A more exotic
Kantowski-Sachs metric was considered in \cite{jensen89}, there
$T=2$ with ${\cal M}_{(1)}=S^1$, ${\cal M}_{(2)}=S^2$.

The equations of motion for the scale factors are derived using
the Einstein-Hilbert action, for which we need to calculate the
components of the Riemann tensor. This is made simpler by using
methods from differential geometry \cite{eguchi80}, so we start by
defining an orthonormal basis of one forms on ${\cal M}$.
\begin{eqnarray}
\label{orth_basis}
\omega^0&=&{\rm d}\xi\\
\nonumber
\omega^{\bar{\mu}}_{(i)}&=&a_{(i)}\bar{\omega}^{\bar{\mu}}_{(i)}
\end{eqnarray}
The notation we are using is that barred quantities correspond to
properties on the submanifolds ${\cal M}_{(i)}$. So, in
(\ref{orth_basis}) the $\bar{\omega}^{\bar{\mu}}_{(i)}$ are an
orthonormal basis of one forms with respect to the metric
$\rm{ds}_{(i)}^2$ and \mbox{$\bar{\mu}=1...n_{(i)}$}, whereas
the \mbox{$\omega^{\bar{\mu}}_{(i)}$} are in the orthonormal basis of
$\rm{ds}^2$. The notation
for the orthonormal basis of $\rm{ds}^2$ ($\omega^\mu$) is such
that \mbox{$\omega^\mu=\omega^{\bar{\mu}}_{(i)}$},
\mbox{$\mu=1...N$}. So because \mbox{$\bar{\mu}=1...n_{(i)}$} we
find \mbox{$\bar{\mu}=\mu-(n_{(1)}+n_{(2)}+...+n_{(i-1)})$}. This
should be unambiguous (although it might not seem so at first
glance!) as a barred index always appears on a quantity with a
subscript $(i)$ saying which ${\cal M}_{(i)}$ it lives on.

The connection one forms on the ${\cal M}_{(i)}$
($\bar{\Theta}^{\bar{\mu}}_{(i)\bar{\nu}}$) are taken to be
torsion free metric connections,
\begin{eqnarray}
\label{Mi_connection}
\rm{d}\bar{\omega}^{\bar{\mu}}_{(i)}
+\bar{\Theta}^{\bar{\mu}}_{(i)\bar{\nu}}
\wedge\bar{\omega}^{\bar{\nu}}_{(i)}&=&0\\
\nonumber
\bar{\Theta}_{(i)\bar{\nu}\bar{\mu}}&=&
-\bar{\Theta}_{(i)\bar{\mu}\bar{\nu}}
\;\;\;\;\;\;\;\;\;\;\bar{\mu},\bar{\nu}=1...n_{(i)}.
\end{eqnarray}
The connection forms on ${\cal M}$ satisfy similar relations
\begin{eqnarray}
\label{M_connection}
\rm{d}\omega^{\mu}
+\Theta^{\mu}_{\;\;\nu}\wedge\omega^{\nu}&=&0\\
\nonumber
\Theta_{\nu\mu}&=&-\Theta_{\mu\nu}
\;\;\;\;\;\;\;\;\;\;\mu,\nu=0...N.
\end{eqnarray}
To evaluate the connection one forms we use (\ref{orth_basis}) to find
\begin{eqnarray}
\rm{d}\omega^0&=&0\\
\rm{d}\omega^{\bar{\mu}}_{(i)}&=&
\alpha'_{(i)}\omega^0\wedge\bar{\omega}^{\bar{\mu}}_{(i)}
                 +a_{(i)}\rm{d}\bar{\omega}^{\bar{\mu}}_{(i)},
\end{eqnarray}
where we have introduced \mbox{$\alpha_{(i)}=\ln(a_{(i)})$} and
the prime denotes the derivative with respect to $\xi$. Taking the
definition of $\Theta^{\mu}_{\;\;\nu}$ in (\ref{M_connection}) and
using (\ref{Mi_connection}) for the individual ${\cal M}_{(i)}$ we
find
\begin{eqnarray}
\label{connection_sol}
\Theta^0_{(i){\bar{\mu}}}&=&-\alpha'_{(i)}\omega^{\bar{\mu}}_{(i)}\\
\Theta^{\bar{\mu}}_{(i){\bar{\nu}}}&=&
\bar{\Theta}^{\bar{\mu}}_{(i){\bar{\nu}}}
\;\;\;\;\;\;\;\;\;\;\bar{\mu},\bar{\nu}=1...n_{(i)}.
\end{eqnarray}

So, if the indices on $\Theta^{\mu}_{\;\;\nu}$ correspond to different
${\cal M}_{(i)}$ then that connection form vanishes.
We must also take care to note that
$\bar{\Theta}^{\bar{\mu}}_{(i){\bar{\nu}}}$ is defined using the basis on
${\cal M}_{(i)}$ ($\bar{\omega}^{\bar{\mu}}_{(i)}$) whereas
$\Theta^{\bar{\mu}}_{(i){\bar{\nu}}}$ uses that on ${\cal M}$
($\omega^{\bar{\mu}}_{(i)}$).

Now that the connection forms on ${\cal M}$ are known, in terms of
those on ${\cal M}_{(i)}$, we may calculate
the curvature two forms \cite{eguchi80},
\begin{eqnarray}
\label{M_curvature_def}
R^\mu_{\;\nu}&=&\rm{d}\Theta^\mu_{\;\nu}
+\Theta^\mu_{\;\rho}\wedge\Theta^\rho_{\;\nu}
\;\;\;\;\;\;\;\;\mu,\nu,\rho=0...N.
\end{eqnarray}
There is an analogous expression for the curvatures on
${\cal M}_{(i)}$, where the appropriate barred connections are used.
Using (\ref{Mi_connection}) one finds for the curvature forms on
${\cal M}$.
\begin{eqnarray}
\label{M_curvature_sol}
R^0_{(i)\bar{\mu}}&=&
  -[\alpha''_{(i)}
+(\alpha'_{(i)})^2]\omega^0\wedge\omega^{\bar{\mu}}_{(i)}
\\
\nonumber
R^{\bar{\mu}}_{(i)\bar{\nu}}&=&\bar{R}^{\bar{\mu}}_{(i)\bar{\nu}}
      -(\alpha'_{(i)})^2\omega^{\bar{\mu}}_{(i)}
       \wedge\omega^{\bar{\nu}}_{(i)}\\
\nonumber
R^{\bar{\mu}}_{(i,j)\bar{\bar{\nu}}}&=&
       -\alpha'_{(i)}\alpha'_{(j)}\omega^{\bar{\mu}}_{(i)}
        \wedge\omega^{\bar{\bar{\nu}}}_{(j)}
\end{eqnarray}
The notation for the last equation of (\ref{M_curvature_sol}) is that
$i\neq j$ and the single barred index corresponds to ${\cal M}_{(i)}$
with the double barred index living on ${\cal M}_{(j)}$. Again we
stress that $\bar{R}^{\bar{\mu}}_{(i)\bar{\nu}}$ is defined with the
$\bar{\omega}^{\bar{\mu}}_{(i)}$ basis, which differs from the basis
on ${\cal M}$ by a factor of $a_{(i)}(\xi)$.

Einstein's equations relate the Ricci tensor to the stress-energy
tensor. For the above curvature two forms we use
\mbox{$R^\mu_{\;\nu}=
\frac{1}{2}{\cal R}^\mu_{\;\nu\rho\sigma}
\omega^\rho\wedge\omega^\sigma$},
enabling us to calculate the Ricci tensor
\mbox{${\cal R}_{\mu\nu}={\cal R}^\rho_{\;\mu\rho\nu}$} and Ricci
scalar \mbox{${\cal R}={\cal R}^\mu_{\;\mu}$}:
\begin{eqnarray}
\label{ricci}
{\cal R}_{00}&=&-n_{(1)}[\alpha''_{(1)}+(\alpha'_{(1)})^2]
                -n_{(2)}[\alpha''_{(2)}+(\alpha'_{(2)})^2]...\\
\label{ricci_a}
{\cal R}^{(i)}_{\bar{\mu}\bar{\nu}}&=&
\frac{1}{a_{(i)}^2} \bar{{\cal R}}^{(i)}_{\bar{\mu}\bar{\nu}}
\;\;\;\;(\mu\neq\nu)\\
\label{ricci_b}
{\cal R}^{(i)}_{\bar{\mu}\bar{\mu}}&=&
    -\alpha''_{(i)}
    +\frac{1}{a_{(i)}^2} \bar{{\cal R}}^{(i)}_{\bar{\mu}\bar{\mu}}
    -\alpha'_{(i)}[n_{(1)}\alpha'_{(1)}+n_{(2)}\alpha'_{(2)}+...]\\
\label{ricci_c}
{\cal R}&=&-2\left(n_{(1)}\alpha''_{(1)}+n_{(2)}\alpha''_{(2)}+...\right)
           -\left(n_{(1)}(\alpha'_{(1)})^2+n_{(2)}(\alpha'_{(2)})^2
           +...\right)\\
\nonumber
           &~&-\left(n_{(1)}\alpha'_{(1)}+n_{(2)}\alpha'_{(2)}+...\right)^2
           +\left( \frac{1}{a_{(1)}^2}\bar{{\cal R}}^{(1)}
                  +\frac{1}{a_{(2)}^2}\bar{{\cal R}}^{(2)}+...\right),
\end{eqnarray}
where the repeated $\bar{\mu}$ index in (\ref{ricci_b}) is not summed over.
The Einstein tensor is defined by
\mbox{$G_{\mu\nu}={\cal R}_{\mu\nu}-\frac{1}{2}g_{\mu\nu}{\cal R}$},
which with the Einstein-Hilbert action (\ref{EHaction}) leads to
\begin{eqnarray}
\label{einstein_eqns}
G_{00}&=&\frac{1}{2}\phi'^2-{\cal V}(\phi)\\
\label{einstein1}
G^{(i)}_{\bar{\mu}\bar{\nu}}&=&0\\
\label{einstein2}
G^{(i)}_{\bar{\mu}\bar{\mu}}&=&-\frac{1}{2}\phi'^2-{\cal V}(\phi)
\;\;\;\;\;\textnormal{(no sum)}.
\end{eqnarray}
The key breakthrough now is to realise that for (\ref{einstein1})
and (\ref{einstein2})
to be consistent then $\bar{{\cal R}}^{(i)}_{\bar{\mu}\bar{\nu}}$
must be constants for \mbox{$\bar{\mu}=\bar{\nu}$}
and vanish otherwise. As we are using an orthonormal basis, this
is precisely the statement that 
the metrics $\rm{ds}^2_{(i)}$ are Einstein metrics.  In fact the
equations are independent of what that metric is because the only
effect a different Einstein metric has is to change the constant
of proportionality between the Ricci and metric tensor, which may 
then be absorbed into the scale factors (if it is non-vanishing).
This means
we may replace $\bar{{\cal R}}^{(i)}_{\bar{\mu}\bar{\mu}}$ in
(\ref{ricci_b}) by \mbox{$\Lambda_{(i)}=0,\pm 1$} and $\bar{{\cal
R}}^{(i)}$ in (\ref{ricci_c}) by $n_{(i)}\Lambda_{(i)}$ without
loss of generality, as long as we remember to rescale the action appropriately.

We may also see what boundary terms are required in
(\ref{EHaction}) by integrating the Ricci scalar by parts to find
the boundary contribution. The volume form on ${\cal M}$ is given
by the wedge product of the volume forms on the ${\cal M}_{(i)}$,
\begin{eqnarray}
\eta&=&a_{(1)}^{n_{(1)}}a_{(2)}^{n_{(2)}}...a_{(T)}^{n_{(T)}}
\omega^0\wedge\eta_{(1)}\wedge\eta_{(2)}...\wedge\eta_{(T)}.
\end{eqnarray}
By defining $V_{(i)}$ as the volume of ${\cal M}_{(i)}$ and
$\beta$ to be the product of the scale factors we find
that the action, including boundary terms, is given by
\begin{eqnarray}
\label{tot_action}
S_{\rm{E}}&=&V_{(1)}V_{(2)}...V_{(T)}\left\{
-\left[\frac{\partial\beta}{\partial \xi}\right]_{\xi_S}^{\xi_{N}}
+\int {\rm d}\xi\beta\left[-\frac{1}{2}{\cal R}
                     +\frac{1}{2}\phi'^2+{\cal V}(\phi)\right]
\right\}\\
\label{eff_action}
&=&V_{(1)}V_{(2)}...V_{(T)}\left\{
-\left[\frac{\partial\beta}{\partial \xi}\right]_{\xi_S}^{\xi_{N}}
-\frac{2}{n_{(1)}+n_{(2)}+...+n_{(T)}-1}\int {\rm d}\xi \beta(\xi){\cal V(\phi)}
\right\}\\
\label{beta}
\beta(\xi)&=&a_{(1)}^{n_{(1)}}a_{(2)}^{n_{(2)}}...a_{(T)}^{n_{(T)}}.
\end{eqnarray}
In arriving at the second equation we have used the trace of
Einstein's equations (\ref{einstein_eqns})-(\ref{einstein2}) to eliminate 
the scalar curvature and scalar field kinetic terms. The quantities
$\xi_S$ and $\xi_N$ refer to the range of the $\xi$ coordinate, with
$\xi_N$ being the `north' pole and $\xi_S$ referring to the `south' 
pole of the instanton taken to be $\xi=0$. To save writing out the
volumes of all the submanifolds we shall call the term in the curly
braces of (\ref{eff_action}) the {\it reduced action}.

The preceding calculation shows that for metric (\ref{M_metric})
to be consistent then the metrics on the ${\cal M}_{(i)}$ must be
Einstein. Given this, the evolution equations for the scale
factors $a_{(i)}(\xi)$ depend only on the value of the
`cosmological constants' $\Lambda_{(i)}$ and not on the detailed
topology or geometry of the manifold. 
This is potentially very significant,
for any statements we can make about the evolution of the scale
factors cover a very large class of manifolds, all those admitting
an Einstein metric. 
Whilst there are manifolds which do not admit an Einstein metric
due to topological restriction, there is a large range which do.
For example, all semi-simple Lie groups have a
Killing metric which is Einstein, along with 
a large class of quotient spaces.
It is noted that any given manifold
may have more than one Einstein metric, \cite{besse}.

\section{Numerical solutions and Actions for Instanton configurations.}

To make some specific predictions we shall in this section
numerically investigate the case where there are
just two submanifolds ${\cal M}_a$, ${\cal M}_b$ with scale
factors $a(\xi)$ and $b(\xi)$ respectively. 
The dimension of and `cosmological constant' associated with
these manifolds are taken as $n_a$, $n_b$ and $\Lambda_a$,
$\Lambda_b$ respectively. The equations of motion are then,
\begin{eqnarray}
\label{ab_constraint}
\frac{1}{2}n_a(n_a-1)\alpha'^2+\frac{1}{2}n_b(n_b-1)\beta'^2
+n_a n_b\alpha'\beta' -\frac{1}{2}\frac{n_a\Lambda_a}{a^2}
-\frac{1}{2}\frac{n_b\Lambda_b}{b^2} &=&\frac{1}{2}\phi'^2-{\cal V}\\
(n_a-1)\frac{a''}{a} + n_b \frac{b''}{b}
+\frac{1}{2}(n_a-2)\left[(n_a-1)\alpha'^2-\frac{\Lambda_a}{a^2}\right]
+\frac{1}{2}n_b\left[(n_b-1)\beta'^2-\frac{\Lambda_b}{b^2}\right]
+n_b(n_a-1)\alpha'\beta'
&=&-\frac{1}{2}\phi'^2-{\cal V}\\
n_a\frac{a''}{a}+(n_b-1) \frac{b''}{b}
+\frac{1}{2}n_a\left[(n_a-1)\alpha'^2-\frac{\Lambda_a}{a^2}\right]
+\frac{1}{2}(n_b-2)\left[(n_b-1)\beta'^2-\frac{\Lambda_b}{b^2}\right]
+n_a(n_b-1)\alpha'\beta'
&=&-\frac{1}{2}\phi'^2-{\cal V}\\
\label{phieqn}
\phi''+(n_a\alpha'+n_b\beta')\phi'&=&\frac{\partial {\cal
V}}{\partial \phi}
\end{eqnarray}
In solving these equations, we used a simple potential, namely 
\mbox{$V(\phi)=\frac{1}{2} \phi^2$}, although our
results do not qualitatively depend on the exact 
shape of the potential.
The main solutions of interest here are the cases where the south
pole is regular and there is a curvature singularity at the north
pole \cite{hawking98a}. Other cases where both the north and south
poles are singular have been studied \cite{bousso98}. We shall not 
concentrate on these cases here as the interesting features we wish to discuss are found in the case where only the north pole is singular.

As we want the south pole to be a smooth endpoint 
this places conditions on the scale factors. In order to avoid
a conical curvature singularity then only one scale factor may vanish
there, with all others approaching a constant. We order the 
${\cal M}_{(i)}$ such that $a_{(1)}$ vanishes at $\xi=0$,
according to
\begin{eqnarray}
\label{lambda_constraint}
a_{(1)}(\xi \rightarrow 0) \rightarrow 
\sqrt{\frac{\Lambda_{(i)}}{n_i -1}}\;\;\xi,
\end{eqnarray}
We see then that we must have $\Lambda_{(1)} > 0$
for everything to be well defined and for the solution to be
non-trivial.

Before we proceed we need to know what effect the singularity
is going to have on the action; in particular, does it
remain finite? To decide this we make the assumption that at the
singularity the potential is not
important, although there are exceptions when exponential
potentials are used \cite{saffin98}. We may then integrate
(\ref{phieqn}) to obtain
\begin{eqnarray}
\phi'(\xi\rightarrow\xi_N)\propto(a^{n_a}b^{n_b})^{-1}.
\end{eqnarray}
Now assume a polynomial behaviour for the scale factors near the singularity of the form
\begin{eqnarray}
\label{alimits}
a(\xi\rightarrow\xi_N)\propto(\xi-\xi_N)^p\;\;\;\;\;
b(\xi\rightarrow\xi_N)\propto(\xi-\xi_N)^q.
\end{eqnarray}

This is consistent with (\ref{ab_constraint}-\ref{phieqn})
so long as $(q,p\leq1)$.
Then the dominant behaviour on the left hand side
of (\ref{ab_constraint}) is $(\xi-\xi_N)^{-2}$, giving
$n_a p + n_b q =1$. The volume factor, $\beta=a^{n_a}b^{n_b}$ 
(\ref{eff_action}), then goes linearly to zero at $\xi_N$.
Moreover, as $\phi'$ is diverging as $(\xi-\xi_N)^{-1}$,
$\phi$ diverges logarithmically which is slow enough that the
linear volume factor renders the singularity integrable for our potential.

We include some representative results below for the case 
of two Einstein metrics of dimensions $n_{(1)}=3$ and 
$n_{(2)}=2$. To be specific we have chosen to take the value
of \mbox{$\Lambda_{(i)}=n_{(i)}-1$}, which is the 
appropriate value for
the round metric on S$^{n_{(i)}}$. We shall explain the reason
for this shortly.

\begin{figure}[h]\centering
\leavevmode\epsfysize=6cm \epsfbox{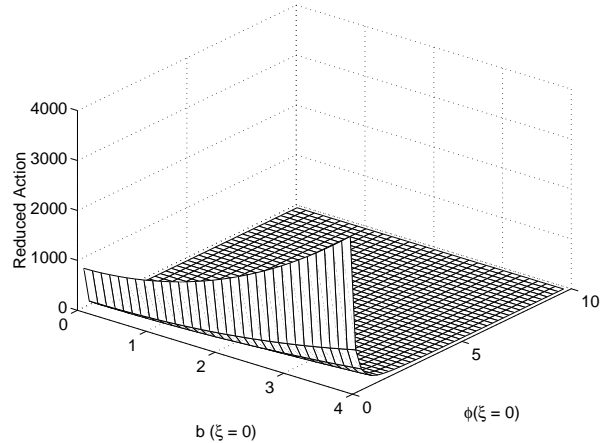}\\
\caption{Plot of the reduced action for case 
$n_{(a)}= 3, n_{(b)} = 2$}
\end{figure}

\begin{figure}[ht]\centering\leavevmode\epsfysize=6cm 
\label{reduced_detail}
\epsfbox{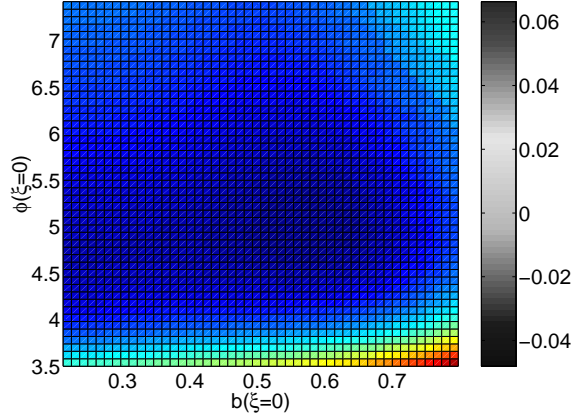}\\
\caption{Detail of the reduced action plot showing a minima for
$n_{(a)}= 3, n_{(b)} = 2$}
\end{figure}

The second of the plots above is a magnification of some 
structure on the first, showing a minima with negative reduced
action. So, just as the Hawking-Turok instanton had a minima
in the action so does this new solution which has non-trivial `spatial' topology. The issue we need to address now is which has a lower
action and is therefore more likely as initial conditions.
For this we need to know the full Euclidean action rather than
the reduced action, this means that the volume prefactors in 
(\ref{eff_action}) must be found. 
Naively it may seem that the total action may
be made arbitrarily large and negative by choosing the Einstein
metrics such that the volume of the manifolds is arbitrarily
large. However, a result from Riemannian geometry, Bishop's
theorem \cite{besse} implies that for positive $\Lambda$ the metric
which maximises the volume of the manifold is the round sphere.
So by looking at the volumes appropriate for the round
metric we are looking
at the lowest possible action for {\it all} 
\mbox{$\Lambda_{(i)}>0$}
instantons. The volume of S$^n$ with the round metric is
\begin{eqnarray}
{\textnormal Vol}(\rm{S}^n)&=&2\pi^{\frac{n+1}{2}}/\Gamma\left(\frac{n+1}{2}\right),
\end{eqnarray}
so for our example of $n_{(1)}=3$, $n_{(2)}=2$ the total action
is \mbox{$S_{\rm{E}}=(2\pi^2)(4\pi)(-0.048)=-11.9$}, where the reduced action
of -0.048 follows from the minima of Fig. 2.
We now need to see how this compares to the Hawking-Turok solution in higher
dimensions.
The starting point is the metric ansatz. This corresponds to 
(\ref{M_metric}) with $T=1$ and $\rm{ds}^2_{(1)}$ the round metric 
on S$^n$, which gives the equations
\begin{eqnarray}
\label{hta}
\frac{a''}{a} &=&- \left( 2\frac{{\cal V}(\phi)}{n(n-1)}
  +\frac{\phi'^2}{n} \right) \\
\label{htphi}
\phi''+n\phi' \frac{a'}{a} &=&  {\cal V}'(\phi).
\end{eqnarray}
The action is:
\begin{equation}
S_E= -{\textnormal Vol}(\rm{S}^n)\left((a(\xi_N)^n)'+ 
\frac{2}{n-1} \int d \xi(a(\xi)^n {\cal V}(\phi))\right).
\label{htact1}
\end{equation}

We may get an approximate solution to these equations by following
the process laid out in \cite{hawking98}.
To start we integrate (\ref{htphi}) to find
\begin{eqnarray}
\left(a(\xi_N)^n \phi'(\xi_N) \right)&=&
\int^{\xi_N}_{0}{\rm d}\xi a(\xi)^n\frac{\partial {\cal V}}{\partial \phi},
\end{eqnarray}
then we make the approximation that at $\xi_N$ the constraint
equation (\ref{ab_constraint}) yields,
\begin{eqnarray}
a'(\xi_N)&\simeq&-\frac{a(\xi_N)\phi'(\xi_N)}{\sqrt{n(n-1)}}
\end{eqnarray}
The action is then found from (\ref{htact1}) by taking the
scalar field to be the constant
\mbox{$\phi_0=\phi(\xi=0)$},
\begin{equation}
\label{htact1b}
S_E\simeq {\textnormal Vol}(\rm{S}^n)\left( -\frac{2{ \cal V}(\phi_0)}{n-1}
    +\left(\frac{n}{(n-1)}\right)^{1/2}
{\cal V},_{\phi_0} \right) \int \rm{d}\xi a(\xi)^n,
\end{equation}
The next approximation is that
\mbox{$a(\xi)\simeq H^{-1}\sin(H\xi)$}, where
\mbox{$H^2= \frac{2{\cal V}(\phi_0)}{n(n-1)}$}. This then leaves us with,
\begin{equation}
S_E\simeq \left( -\frac{2{ \cal V}(\phi_0)}{n-1}
    +\left(\frac{n}{(n-1)}\right)^{1/2}
{\cal V},_{\phi_0} \right) \frac{I_n}{H^{(n+1)}}Vol(\rm{S}^n),
\label{htact2}
\end{equation}
\begin{eqnarray}
 I_n = \left\{ \begin{array}{ll}
      2^n[ ((n-1)/2)! ]^2 \over n!
              &\mbox{if n is odd}\\
      (n-1)!\pi \over 2^{(n-1)}(n/2-1)!(n/2)!
              &\mbox{\rm if~ n~ is~ even}.
              \end{array}
\right.
\end{eqnarray}
For example we find, $I_2=\pi/2,~I_3=4/3,~I_4=3\pi/8,~I_5=16/15$. 
For the
archetypal harmonic potential,  ${\cal V}=\frac12 \phi^2$,
 we obtain an estimate
for the location of  the minima to be
$(\phi_0)_{\rm min}\simeq n\sqrt{\frac{n}{(n-1)}}$,
which is approximately linear in n. The
corresponding value for the action is
\begin{equation}
S_{\rm min} (n) \simeq -I_n \left[\frac{n-1}{n} \right]^{(n-1)}Vol(\rm{S}^n).
\end{equation}
The full numerical solutions to (\ref{hta})-(\ref{htact1}) are given in
Fig. 3, where the reduced action is found for a 
range of dimensions.
The behaviour is well explained by the approximation scheme, which
describes the positions of the minima increasing as $n$ increases,
along with the value of the reduced action which also increases with n.

\begin{figure}[ht]
\label{htplot}
\centering\leavevmode\epsfysize=6cm 
\epsfbox{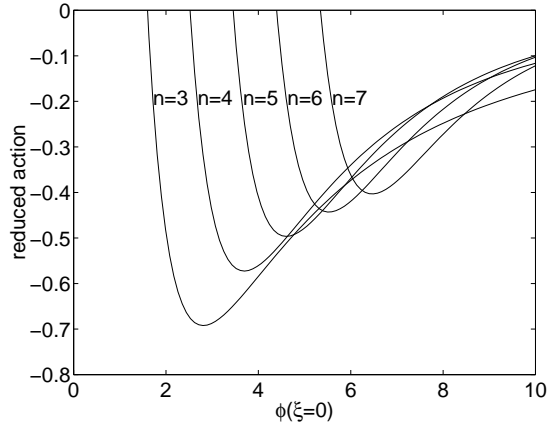}\\
\caption{Numerical results for Hawking Turok instantons of different 
dimensionalities, 
\mbox{$\rm{ds}^2={\rm d}\xi^2+a(\xi)^2{\rm d}\Omega^2_n$}}
\end{figure}

We are now in a position to compare the total action we found for the
product ansatz, -11.9, and this more symmetric case. The example we found
before, Fig. 2, was for a six dimensional manifold.
the total action of  the Hawking-Turok solution in six dimensions
is seen from Fig. 3 as $\pi^3(-0.495)=-15.3$, which is lower.
We have checked this for a number of dimensions, using
\mbox{$\Lambda_a$, $\Lambda_b>0$}, and found that
although a minima existed it had a higher action than the
corresponding Hawking-Turok instanton.
Although this is not a proof that the lowest
action will not be of the product form, we see no reason to suspect
otherwise.

Now let us proceed to investigate 
analytically the behaviour of the action
for a more general class of our instantons to see if we can
understand why a minima appears at all. Consider eqn 
(\ref{ab_constraint}).
If we take p and q of (\ref{alimits}) to both be less than one then, 
using our knowledge of the
asymptotic behaviour of the field and scale factors, we find that near a
singular pole this equation becomes,

\begin{eqnarray}
\label{approx1}
\frac{1}{2}n_a(n_a-1)\alpha'^2+\frac{1}{2}n_b(n_b-1)\beta'^2
+n_a n_b\alpha'\beta' =\frac{1}{2}\phi'^2.
\end{eqnarray}

Let us consider the case $n_a, n_b >> 1$. Then if we multiply eqn.
(\ref{approx1}) through by $a^{2n_a}b^{2n_b}$ we find that the left
hand side becomes approximately equal to the boundary term in 
the action squared.
Using this observation we can once more follow through the 
analysis of Hawking and
Turok but this time the analysis will apply to the case of 2 
submanifolds.
The analysis here is the same as that which lead to (\ref{htact1b}),
except that we must take \mbox{$n_a$, $n_b>>1$} in the constraint
equation to find
\begin{eqnarray}
S_{E} = \left({\cal V},_{\phi_{0}} 
- \frac{2{\cal V}(\phi_{0})}{(n_a + n_b -1)}\right) V_{(1)} V_{(2)}
\int \rm{d}\xi a(\xi)^{n_a}b(\xi)^{n_b}
\end{eqnarray}
The next step is to make the approximation that
\mbox{$a(\xi)\simeq H^{-1}\sin(H\xi)$} and
\mbox{$b(\xi)=b(\xi=0)=$constant}. The end result is that the action
is approximated by the value obtained in (\ref{htact2}) multiplied
by $b(\xi=0)^{n_b}$ and an extra submanifold volume factor.
This then explains why a valley is found in Fig. 2 which is parallel
to the $b(0)$ axis, getting deeper as $b(0)$ increases. Clearly the
valley cannot keep getting arbitrarily deep, so at some point these
approximations break down. We find that the weak link in our chain
of reasoning is the assumption that we make about 
the behaviour of the scale factors as they approach the north pole.
For small $b(0)$ then we find that $a(\xi)$ decreases to zero at
$\xi_N$, and our approximation works. When $b(0)$ is larger than
some critical value then instead of decreasing to zero,
$a(\xi)$ increases and diverges at $\xi_N$ and  the approximation
of treating it as a sine breaks down.

The important point to take away from all this is that some of our
types of instanton with a `warp product space' topology have local 
minima of action in parameter space, 
but the corresponding Hawking Turok
instanton (with appropriate dimensionality) will still dominate over
them in the semi-classical approximation to the Hartle Hawking
wavefunction.

\subsection{A conjecture.}

We have seen that the metric associated with the scale factor which
vanishes at $\xi_S$ must have \mbox{$\Lambda_1>0$}, 
(\ref{lambda_constraint}).
The same constraint does not apply to the other manifolds. If we
were to allow \mbox{$\Lambda_{i>1}\leq 0$} then Bishop's theorem
does not put a limit on the total 
volume of the ${\cal M}_{(i>1)}$; so if the
reduced action had a minima with a negative value then 
(\ref{eff_action}) could be made arbitrarily negative by increasing
$V_{(i>1)}$. We would therefore expect that 
{\it negative values of the reduced action occur only if}
$\Lambda_{(i)}>0$ {\it for all} $i$.
We have checked this for the case of two submanifolds of various
dimensions, always confirming this conjecture.

\section{Analytical solutions and analytical continuation.}

The equations are simplified considerably if we actually drop the
scalar field $\phi$, and replace its potential with a cosmological
constant $\Lambda$. We then obtain the following analytical
solutions \cite{garriga98}. The first is given by,

\begin{eqnarray}
a_{(1)} = \frac{1}{ \sqrt{(n_{(1)}-1)}\;\chi}sin(\chi\xi)
\end{eqnarray}
\begin{eqnarray}
n_{(1)}(n_{(1)}-1)\chi^2 = 2 \Lambda - \sum_{i > 1}( \frac{n_{(i)} 
\Lambda_{(i)}}{a_{(i)}^2})
\end{eqnarray}
and for $i>1$,
\begin{eqnarray}
\label{const_scal}
a_{(i)} = \sqrt{ \frac{ \Lambda_{(i)} }{ n_{(1)} \chi^2 }},
\end{eqnarray}
where $n_{(1)}>1$. There is a similar solution when $n_{(1)}=1$. 
It is also possible that $\chi$ may be taken as imaginary, in which
case the trigonometric function become hyperbolic, rendering
the instanton non-compact.
One can still create finite size instantons in this case
by introducing domain walls at some value $\xi_W$. This creates
a discontinuity in the gradient of $a_{(1)}$ causing the scale
factor to decrease, if the wall tension is large enough.
We can see that the limiting behaviour
\mbox{$a_{(1)}(\xi\rightarrow 0)
\rightarrow 1/\sqrt{(n_{(1)}-1)}$} is consistent with
(\ref{lambda_constraint}) for \mbox{$\Lambda_{(1)}=1$}, 
meaning that the metric is regular at the end points.
Equation (\ref{const_scal}) also shows that if we have $\chi$
real then \mbox{$\Lambda_{(i>1)}>0$},
and for imaginary $\chi$ \mbox{$\Lambda_{(i>1)}<0$}.
Ricci flat submanifolds would mean a vanishing scale factor
for that manifold, so in effect they are not present. 

The second analytical solution is,

\begin{eqnarray}
a_{(1)} = \frac{1}{ \sqrt{(n_{(1)}-1)}\chi}sin(\chi \xi)
\end{eqnarray}
\begin{eqnarray}
a_{(2)} = \frac{1}{ \sqrt{(n_{(2)}-1)}\chi}cos(\chi \xi)
\end{eqnarray}
\begin{eqnarray}
(2n_{(1)}n_{(2)} + n_{(1)}(n_{(1)}-1) +n_{(2)}(n_{(2)}-1))\chi^2 =
2 \Lambda - \sum_{i > 2}( \frac{n_{(i)}
\Lambda_{(i)}}{a_{(i)}^2})
\end{eqnarray}
and for $i>2$,
\begin{eqnarray}
a_{(i)} = \sqrt{ \frac{ \Lambda_{(i)} }{ (n_{(1)} + n_{(2)})\chi^2}},
\end{eqnarray}
where $n_{(1)}>1$ and $n_{(2)}>1$ although there is a similar
solution when they both are equal to one. 
As before, if $\chi$ is imaginary we require a domain wall to make the
instanton compact. This solution also requires $\Lambda_{(1)} = 1$ and
$\Lambda_{(2)} = 1$ if the instanton is to close off in a regular
manner at the `north' and `south' poles. 

Although the majority of the instantons considered in this paper do
not analytically continue to lorentzian
space times where some of the dimensions are compactified we can use
these analytical solutions to demonstrate
that there are some that do. One subset of the solutions given above is the
product of S$^4$ with S$^n$. This can be analytically 
continued to a four dimensional deSitter space with a static S$^n$ as
the internal manifold.

It should be noted that the static nature 
of this internal manifold is not stable 
to perturbations. This is of course the 
manifestation in this context of the problem 
of stabilising moduli fields in cosmology. 
We do not attempt to resolve this difficulty 
here.
Some recent mechanisms for stabilising moduli fields 
in cosmology can be found in \cite{tiago} 
and \cite{huey}.

\section{Conclusions}

In this paper, we have derived the equations of motion for a
specific warp product of general Einstein metrics. The main conclusion we
can draw is that instantons which continue to spaces with compact
`extra' dimensions of the form considered here do not have lower action
than the corresponding higher-dimensional Hawking-Turok instanton.
However, non trivial minima of the action do occur if the Einstein
metrics {\it all} have positive $\Lambda_i$.
These results are significant: First, they seem to indicate that the symmetry
arguments used by Hawking and Turok in their letter can be applied
to higher dimensional cases. Secondly, our analysis applies to a
wide range of metrics and cosmological scenarios. Our particular
comparison of the instantons involved n-dimensional spheres as
our internal compact dimensions. Bishop's theorem then implies
that these will
provide the lowest possible action for such
spacetimes with compact internal dimensions, hence our results
apply to any Einstein metric -- they will always be beaten by the
corresponding Hawking-Turok instantons. This result strongly suggests to
us that if the initial quantum state of the universe were to be described by the
`Hartle Hawking proposal' then it would be difficult to explain the
presence of extra compact dimensions.


\section*{Acknowledgements}

We are grateful to J. Garriga, S. Gratton, N. Manton and 
T. Wiseman for useful
conversations. EJC, JG and PS are all supported by PPARC. The
numerical work was carried out in the UK-CCC COSMOS Origin 2000
supercomputer, supported by Silicon Graphics/Cray Research, HEFCE
and PPARC.



\end{document}